\documentclass[reprint,amsmath,amssymb,floatfix,aps,superscriptaddress]{revtex4-1}

\usepackage{graphicx}
\graphicspath{ {./_Figures/} }
\usepackage{epsfig}
\usepackage{epstopdf}
\usepackage{color}         			% color text
\usepackage[normalem]{ulem}			% Allow strikeout formatting of text
\usepackage{soul} 						 	% strikeout formatting

\begin{document}
		
\title{Resonant Elastic X-ray Scattering from the Skyrmion Lattice in Cu$_{2}$OSeO$_{3}$}
		
\author{S. L. Zhang}
\address{Clarendon Laboratory, Department of Physics, University of Oxford, Parks Road, Oxford, OX1~3PU, UK}

\author{A. Bauer}
\address{TU M\"unchen, Physik-Department E21, D-85748 Garching, Germany}

\author{H. Berger}
\address{Crystal Growth Facility, Ecole Polytechnique F\'{e}d\'{e}rale de Lausanne (EPFL), CH-1015 Lausanne, Switzerland}

\author{C. Pfleiderer}
\address{TU M\"unchen, Physik-Department E21, D-85748 Garching, Germany}

\author{G. \surname{van der Laan}}
\address{Magnetic Spectroscopy Group, Diamond Light Source,	Didcot, OX11~0DE, UK}

\author{T. Hesjedal}
\address{Clarendon Laboratory, Department of Physics, University of Oxford, Parks Road, Oxford, OX1~3PU, UK}

\date{\today}
	
\begin{abstract}
We report the study of the skyrmion state near the surface of Cu$_2$OSeO$_3$ using soft resonant elastic x-ray scattering (REXS) at the Cu $L_3$ edge. Within the lateral sampling area of $200 \times 200$~$\mu$m$^2$, we found a long-range-ordered skyrmion lattice phase as well as the formation of skyrmion domains via the multiple splitting of the diffraction spots. In a recent REXS study of the skyrmion phase of Cu$_2$OSeO$_3$ [Phys.\ Rev.\ Lett.\ \textbf{112}, 167202 (2014)], Langner \textit{et al.}\ reported the observation of the unexpected existence of two distinct skyrmion sublattices that arise from inequivalent Cu sites, and that the rotation and superposition of the two periodic structures leads to a moir\'{e} pattern. However, we find no energy splitting of the Cu peak in x-ray absorption measurements and, instead, discuss alternative origins of the peak splitting.
In particular, we find that for magnetic field directions deviating from the major cubic axes, a multidomain skyrmion lattice state is obtained, which consistently explains the splitting of the magnetic spots into two---and more---peaks.
\end{abstract}

%%%%%%%%%%%%%%%%%%%%%%%%%%%%%%%%%%%%%%%%%%%%%%%%%%%%%%%%%%%%%%

% \pacs{68.65.Ac, 75.60.Ch, 75.30.Gw, 78.70.Dm}

%%   ****** Do not delete this list. ********************
%%   ***  75.30.Hx     Magnetic impurity interaction  **
%%   ***  78.70.Dm    X-ray absorption                        **
%%   ***  75.50.Pp     Magnetic semiconductors        **
%%   ***  75.70.Cn    Magnetic properties of interfaces (multilayers, superlattices, heterostructures)
%%   ***  75.30.Gw   Magnetic anisotropy
%%   ***  73.40.Ns    Metal-non metal contacts
%%   ***  73.20.At      Surface states, band structure, electron density of state
%%   ***  73.61.Ng      Insulators (what many use for TIS)
%%   ***  61.66.Fn      Inorganic compounds (what many use for TIS)
%%   ****************************************************

%%%%%%%%%%%%%%%%%%%%%%%%%%%%%%%%%%%%%%%%%%%%%%%%%%%%%%%%%%%%%%%%%%%%%%%%%%%%
\maketitle

%%%%%%%%%%%%%%%%%%%%%%
%%%%%%%%%%%%%%%%%%%%%%
%%%%%%%%%%%%%%%%%%%%%%%%%%%%%%%%%%%%%%%%%%%%%%%%%%%%%%%%%%%%%%%%%%%%%%%%%%%%%%%%%%%%%%%
\section{Introduction}

Magnetic skyrmions are swirls in a magnetic spin system, analogous to the skyrmion particle originally described in the context of pion fields~\cite{Pf_MnSi_Science_09,2010:Jonietz:Science,2010:Yu:Nature,Tokura_CuOSeO_LTEM_Science_12,Rocsh_MnSi_emergent_NatPhys,2013:Milde:Science,2013:Fert:NatureNano,Tokura_review_skyrmion_Natnano_13,Romming2013,Tokura_CuOSeO-MnSi_ratchet_Natmater_14,2015:Schwarze:NatureMater}. 
Due to their unique topological properties, they are proposed as a promising candidate for the advanced spintronics applications \cite{Tokura_review_skyrmion_Natnano_13}.
The most famous skyrmion-carrying materials system are the helimagnets with the crystalline space group $P2_13$, such as MnSi, FeGe, and Cu$_2$OSeO$_3$ \cite{Pf_MnSi_Science_09,2010:Munzer:PhysRevB, Adams_Domains_2010,2010:Yu:Nature,2011:Yu:NatureMater,Seki2012}. The magnetic phases and formation of the skyrmion lattice phase is well-described by the Ginzburg-Landau equation that takes into account thermal fluctuations \cite{Pf_MnSi_Science_09}. The size of a skyrmion is usually 20-70~nm for these materials, which to a large degree limits the available techniques that can fully characterize their magnetic structure. So far, small angle neutron scattering (SANS) \cite{Pf_MnSi_Science_09} and Lorentz transmission electron microscopy (LTEM) \cite{2010:Yu:Nature} have successfully been applied to characterize the skyrmion lattice phase on a microscopic and macroscopic scale, respectively. On the other hand, spontaneous symmetry breaking and domain formation are the natural consequences of magnetism, suggesting that a similar domain effect may exist in the skyrmion lattice phase. However, this requires a characterization technique that probes the material on a mesoscopic scale, in-between the local probing of LTEM and the macroscopic averaging of neutron diffraction. Here, we present resonant soft x-ray scattering on single crystal Cu$_2$OSeO$_3$, which covers the length scale needed to observe skyrmion domains, revisiting earlier work by Langner \textit{et al.}~\cite{Langner2014}.

Among all $P2_13$-type skyrmion-carrying systems, Cu$_2$OSeO$_3$ is a unique compound due to its complex crystalline structure compared with B20 helimagnets, as well as its dielectric and ferroelectric properties \cite{Seki2012}. It is composed of a complex arrangement of distorted CuO$_5$ square-based pyramids and trigonal bipyramids, and a lone-pair tetrahedral SeO$_3$ unit \cite{Berger_Cu2OSeO3_infrared_PRB_10}. The oxygen atoms in the unit cell are shared among these basic elements. All copper ions possess a divalent oxidation state, however, they are distinguished depending on their oxygen environment.

Below $T_c \approx 60$~K the material displays ferrimagnetic ordering. Bos \textit{et al.}\ \cite{Bos2008} determined the magnetic properties and found an effective moment of $\sim$1.36~$\mu_\mathrm{B}$/Cu, which is lower than the value of $\sim$1.73~$\mu_\mathrm{B}$/Cu expected for Cu$^{2+}$, where only the spin moment plays a role. Such a reduced moment is commonly found in metal oxides. The field dependence measurements at 5~K give a saturation value of 0.5~$\mu_\mathrm{B}$/Cu, i.e., half the value expected for a $S=1/2$ spin system, indicative of a collinear ferrimagnetic alignment. The anti-aligned spins are situated on the two chemically distinct copper sites in a ratio of 3:1. The Cu sites have a strong Dzyaloshinskii-Moriya interaction, which is at the origin of the ferroelectricity of this material \cite{Yang2012}.

Cu$_2$OSeO$_3$ has an energy hierarchy similar to other metallic helimagnets \cite{Tokura_review_skyrmion_Natnano_13}. For the skyrmion phase, the helix propagation orientation is weakly pinned by the cubic anisotropy, and can be easily unpinned by introducing fluctuations, such as an electrical field \cite{White_Cu2OSeO3_SANS_E_field_rot_PRL_14} or a thermal gradient \cite{Tokura_CuOSeO-MnSi_ratchet_Natmater_14}. Density-functional-theory calculations show that the propagation wave vectors along all orientations are degenerate for the skyrmion phase \cite{Fudan_Cu2OSeO3_theory_RPL_12}. Therefore, it is expected that under weak perturbation condition, multiple skyrmion domains can exist. The domains have identical absolute values of the propagation vectors, but differ in orientation. This has been observed in both MnSi and Cu$_2$OSeO$_3$ systems by real-time LTEM, and in Fe$_{1-x}$Co$_x$Si by SANS \cite{2010:Munzer:PhysRevB,Adams_Domains_2010}. One can observe the rotating skyrmion domains, confirmed by the two sets of six-fold symmetric spots of the Fourier transform images \cite{Tokura_CuOSeO-MnSi_ratchet_Natmater_14}.

In a recent study, Langner \textit{et al.}~\cite{Langner2014} reported resonant soft x-ray scattering (REXS) of Cu$_2$OSeO$_3$. In their experiment the wavelength of the polarized x-rays was tuned to the Cu $L_3$ edge, and the magnetic diffraction spots were captured on the CCD camera plane in the (001) Bragg condition. In the camera image, the shape of the magnetic satellites is field- and photon energy-dependent, and develops a fine structure, ultimately splitting into more than one spot. The authors interpret this spot splitting as arising from the moir\'{e} pattern of two superposed skyrmion sublattices, which originate from two inequivalent Cu sites, as evidenced by a 2~eV split in their x-ray absorption spectra.

We performed resonant soft x-ray diffraction experiments on a well-characterized Cu$_2$OSeO$_3$ single crystal \cite{EPFL-ARTICLE-196610} and obtained a reciprocal space maps in the $hk$-plane of the skyrmion phase. In the following we give a detailed description of these measurements and present a critical discussion of Langner \textit{et al.}'s results together with an alternative explanation for the peak splitting of the magnetic diffraction peaks observed in Cu$_2$OSeO$_3$ based on the formation of a multidomain state.

%%%%%%%%%%%%%%%%%%%%%%
%%%%%%%%%%%%%%%%%%%%%%
%%%%%%%%%%%%%%%%%%%%%%%%%%%%%%%%%%%%%%%%%%%%%%%%%%%%%%%%%%%%%%%%%%%%%%%%%%%%%%%%%%%%%%%
\section{Resonant scattering\label{sec:theo}} 

For resonant scattering at the $L_{2,3}$ edge of $3d$ transition metals it is sufficient to take only electric-dipole transitions into account \cite{Gerrit_REXS_Physique_08}.
Further, there are two main characteristics of $3d$ materials worth noting.
First, the photon energy falls into the range of 0.4-1~keV, leading to relatively long wavelengths, which limits the number of accessible materials  for experiments in reflection geometry. At the Cu $L_3$ edge, the x-ray wavelength $\lambda$ is $13.3$~\AA. For B20 helimagnetic metals, such as MnSi, Fe$_x$Co$_{1-x}$Si, or FeGe, the (structural) lattice constant $d$ is around 4.5-4.7~\AA\ \cite{Tokura_review_skyrmion_Natnano_13}. Since $\lambda = 2d \sin \theta$, where $\theta$ is the Bragg angle, no structural Bragg reflection is accessible. Cu$_2$OSeO$_3$, on the other hand, has a relatively large lattice constant ($d$ = 8.925~\AA) \cite{Bos2008} so that the (forbidden) (001) peak is accessible.
This provides an ideal condition for performing resonant elastic x-ray scattering (REXS) in reflection geometry. 
Second, the penetration depth of soft x-rays varies strongly for photon energies across the absorption edge \cite{Gerrit_REXS_Physique_08}. The scattering is more bulk-sensitive for photon energies below the $L_3$ and further above the $L_2$ edges, whereas it is more surface-sensitive at resonance. In Cu$_2$OSeO$_3$ the x-ray attenuation length at normal incidence at the $L_3$ edge maximum is 95~nm, while below the absorption edge the attenuation length is 394~nm \cite{vanderFigueroa}.

In a REXS experiment, magnetic diffraction occurs around a structural Bragg peak, as the local magnetic moments are connected to the magnetic atoms that give rise to the resonant diffraction. Therefore, for scattering from single-crystalline Cu$_2$OSeO$_3$, the positions of the magnetic satellites around (001) yield the information about the orientation and periodicity of the modulation.
The diffraction intensity reveals information about the detailed magnetization configuration $\mathbf{M}(\mathbf{q})$.
This is the basic principle for characterizing magnetic structures and for distinguishing between different magnetic phases.
Note that for space group $P2_13$ the (001) peak is crystallographically forbidden and absent in off-resonant  scattering, however, the anisotropic third-rank tensor stemming from the mixed dipole-quadrupole term allows for the extinction peak to appear for non-centrosymmetric crystals at the x-ray resonance condition \cite{Templeton_PRB_94,Dmitrienko_PRL2012}.

\begin{figure}[ht!]
	\begin{center}
		\includegraphics[width=8.6cm]{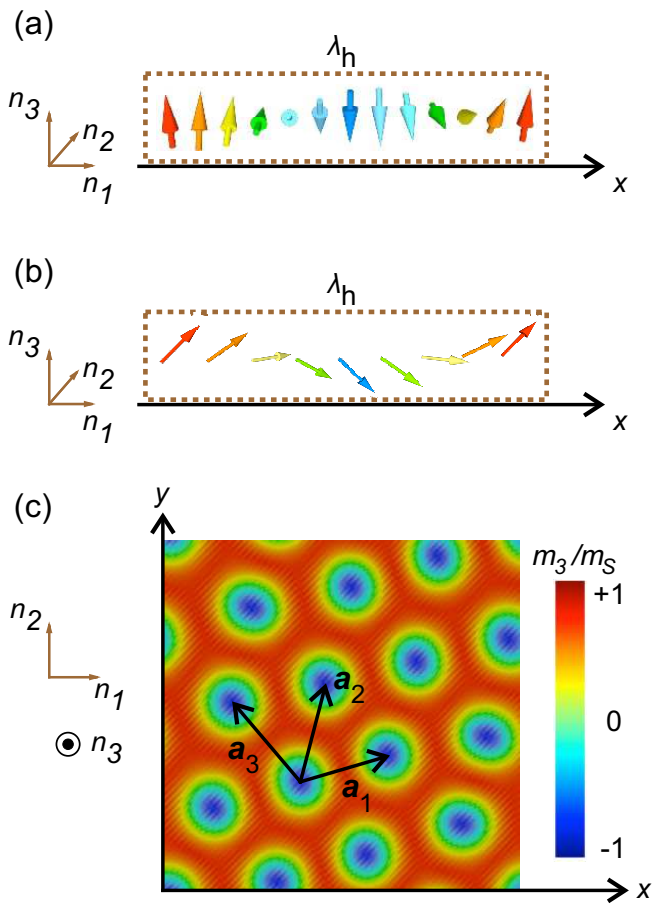}
		\caption{Magnetic elemental units and unit cells of the magnetically ordered phases of Cu$_2$OSeO$_3$. The magnetic elemental units are shown for  (a) helical and (b) conical phase. The magnetic elemental unit for the skyrmion phase is indicated in (c) along with the basis vectors of the magnetic unit cell. The elemental units give rise to the form factors for resonant diffraction and the magnetic unit cells to the structure factors of resonant scattering, as well as the magnetic reciprocal space maps.}
		\label{fig_1}
	\end{center}
\end{figure}

Cu$_2$OSeO$_3$ is a chiral magnet, suggesting ordered spin helices, which further host, in a pocket of the $T$-$H$ (temperature vs magnetic-field) phase space, `crystalline' magnetic order in the form of the skyrmion lattice.
The magnetic modulation is incommensurate, which means that it is decoupled from the atomic lattice. Moreover, the modulation has a long periodicity \cite{Pf_CuOSeO_PRL_12}.
Therefore, the continuum approximation can be applied to model the magnetic properties \cite{Tokura_review_skyrmion_Natnano_13}. 
The system's ground state ($H =0$) is the one-dimensional, helically ordered state composed of single-harmonic modes, where $\mathbf{q}_h$ is the wave vector of the helix with  $\lambda_h=2\pi/{q}_h$ being the real-space helical pitch. The orientation of the modulation is pinned along a $\langle100\rangle$ direction by the cubic anisotropy \cite{White_Cu2OSeO3_E_rotation_PRL_14}.
The magnetization configuration for one helical pitch is illustrated in Fig.\ \ref{fig_1}(a), in which the modulation is along $x$. This is the elemental unit of the helical periodic structure, i.e., the motif of the magnetic crystal.
It is worth mentioning that the helical pitch $\lambda_h$ is equal to the helix-to-helix distance $a_h$ for Cu$_2$OSeO$_3$, as well as other B20 metallic helimagnets. This is well-established by both SANS \cite{Pf_CuOSeO_PRL_12,Tokura_CuOSeO_rotation_PRB_12} and LTEM \cite{Tokura_CuOSeO_LTEM_Science_12,White_Cu2OSeO3_LTEM_PNAS_15} studies.

%%%%%%%%%%%%%%%%%%%%%%
%%%%%%%%%%%%%%%%%%%%%%

Above a certain magnetic field $H_{c1}$, the conical spiral state becomes the lowest energy solution. The single-harmonic spiral has the same pitch as the helical state. The spiral rotates in the $\mathbf{n}_2$-$\mathbf{n}_3$ plane. The magnetization component along $\mathbf{n}_1$-direction is proportional to the magnetic field. At a certain field, $H_{c2}$, all the magnetization vectors are parallel, forming the ferrimagnetic state.
The conical periodic order can also be described by a unit cell that consists of two conical spirals.

The skyrmion vortex structure is a metastable solution in the phenomenological model, unless thermal fluctuation are being taken into account \cite{Pf_MnSi_Science_09}. 
A single skyrmion vortex, as shown in Fig.\ \ref{fig_1}(c), can be written in the form of an axially symmetric magnetization distribution \cite{VA},
\begin{equation}
\begin{aligned}
m_1^\mathrm{sky} (\rho,\phi) &= M_S \sin [\theta(\rho)]\cos [\kappa(\phi+\phi_0)] \,\,\,,\\
m_2^\mathrm{sky} (\rho,\phi) &= M_S \sin [\theta(\rho)]\sin[\kappa(\phi+\phi_0)] \,\,\,,\\
m_3^\mathrm{sky}(\rho,\phi) &= M_S \lambda \cos [\theta(\rho)] \,\,\,\,\,\,\,\,\,\,\,\,\,\,\,\,\,\,\,\,\,\,\,\,\,\,\,\,\,\,\,\,\,\,\,\,,
\label{eq_27}
\end{aligned}
\end{equation}

\noindent using polar coordinates with $\rho=\sqrt{(x^2+y^2)}$ and $\phi=\arctan (y/x)$. $\theta(\rho)$ satisfies the Euler equation and $\kappa$ is the winding number.

Thus, the form factor for an individual skyrmion can be written in the form of
\begin{equation}
f_\mathrm{sky}=\mathbf{V} \iint_{\mathrm{sky}}(\mathbf{\epsilon}_s^*\times\mathbf{\epsilon}_i) (m_1^\mathrm{sky}\mathbf{n}_1 + m_2^\mathrm{sky} \mathbf{n}_2 + m_3^\mathrm{sky}\mathbf{n}_3) e^{i\mathbf{q} \cdot \mathbf{r}} d\mathbf{r}.
\label{eq_29}
\end{equation}

\noindent The integral is taken over the circular area of a skyrmion vortex.
In contrast to the helical and conical states, the skyrmion state is a two-dimensional solution. The `crystal' structure is essentially a hexagonal-type two-dimensional lattice. Therefore, the two-dimensional unit cell can be chosen as shown in Fig.\ \ref{fig_1}(c). The structure factor then becomes
\begin{equation}
F_\mathrm{sky} = f_\mathrm{sky} (1 + e^{i\mathbf{q}\cdot\mathbf{a}_1}+e^{i\mathbf{q}\cdot\mathbf{a}_2}+e^{i\mathbf{q}\cdot\mathbf{a}_3}) \,\,\,\,,
\label{eq_30}
\end{equation}

\noindent where $\mathbf{a}_1$, $\mathbf{a}_2$, $\mathbf{a}_3$ are the real-space basis vectors, which are rotated by $60^\circ$ with respect to each other. The core-to-core distance is $a_1=a_2=a_3$, which can be regarded as the `lattice constant' of the skyrmion crystal.    

%%%%%%%%%%%%%%%%%%%%%%%%%%
\begin{figure}[t!]
	\begin{center}
		\includegraphics[width=8.6cm]{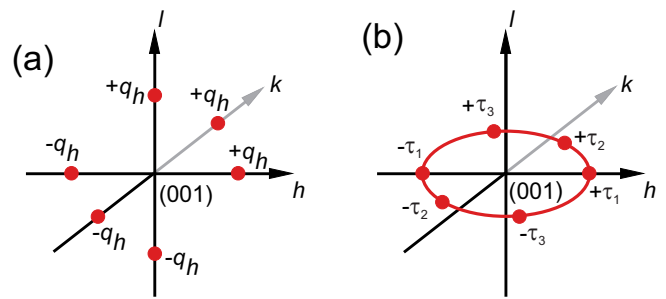}
		\caption{(a) The helical phase is obtained by repeatedly stacking 1D helical chains to form a 2D magnetization pattern. The corresponding magnetic satellites come in pairs of $\pm {\mathbf{q}}_h$, lying on perpendicular axes. (b) Magnetic satellites in reciprocal space for the skyrmion state, where the spots are in the $hk$ plane ($k$ is perpendicular to the plane of the paper).}
		\label{fig_2}
	\end{center}
\end{figure}
%%%%%%%%%%%%%%%%%%%%%%%%%%

Using the form factors of the helical ($f_\mathrm{h}$), conical ($f_\mathrm{c}$), and skyrmion ($f_\mathrm{sky}$) motifs, as well as their structure factors $F_\mathrm{h}$, $F_\mathrm{c}$, and $F_\mathrm{sky}$ of the unit cells, one can easily obtain the reciprocal space maps. In the helical state, the `lattice constant' is equal to the helical pitch. Therefore, the first-order diffraction peaks appear at $\pm\mathbf{q}_h$, and the reciprocal lattice is purely one-dimensional. In Cu$_2$OSeO$_3$, the direction of $\mathbf{q}_h$ is not entirely degenerate, but additionally governed by the sixth order magnetic anisotropy, giving rise to the three-fold degenerate preferred orientation along the three equivalent $\langle 001 \rangle$ directions. Consequently, three spatially separated helical domains are expected. Moreover, the helical magnetic reciprocal space lattice has to be imposed on the crystalline reciprocal space lattice in order to obtain the diffraction condition. The reciprocal space of the helical crystal is plotted in Fig.\ \ref{fig_2}(a), and summarized in Table \ref{tab_1}.
In the conical state, the reciprocal space is similar to the one of the helical phase in that the first order diffraction peaks (modulation vector) appear at $\pm\mathbf{q}_h$. On the other hand, the direction of $\pm\mathbf{q}_h$ is entirely governed by the magnetic field direction, and in fact parallel to it. Therefore, there is only a `single domain' state observed, as summarized in Table \ref{tab_1}.
In the skyrmion state, the reciprocal space [cf., Fig.\ \ref{fig_2}(b)] has three reciprocal-space basis vectors $\mathbf{\tau}_1$, $\mathbf{\tau}_2$, and $\mathbf{\tau}_3$. They are separated by $60^\circ$ and are related to the three lattice constants by $a _i= 2 \pi / (\sqrt{3}\tau_i)$. Therefore, the diffraction peaks appear at $\pm \mathbf{\tau}_i$ around the (001) diffraction, as shown in Fig.\ \ref{fig_2}(b) and   Table \ref{tab_1}.

%%%%%%%%%%%%%%%%%%%%%%%%%%
\begin{table}[htb]
	\caption{Magnetic modulation vectors for the magnetic phases of Cu$_2$OSeO$_3$, and the associated magnetic satellites observable in a REXS experiment.}
	\label{tab_1}
	\begin{tabular}{lcc}
		\toprule
		Phase &  Modulation vectors & Magnetic reflections \\
		\hline
		%\midrule
		Helical & (0,0,$q_h$), (0,0,-$q_h$) & (0,0,1$\pm q_h$)\\
		& ($q_h$,0,0), (-$q_h$,0,0) & ($\pm q_h$,0,1) \\
		& (0,$q_h$,0), (0,-$q_h$,0) & (0,$\pm q_h$,1) \\
		& & \\
		Conical & (0,0,$q_h$), (0,0,-$q_h$) & (0,0,1$\pm q_h$)\\
		& & \\
		Skyrmion & ($\tau$,0,0), (-$\tau$,0,0) & ($\pm \tau$,0,1) \\
		& $\left( -\frac{1}{2}\tau , \frac{\sqrt{3}}{2}\tau ,0 \right)$, $\left( \frac{1}{2}\tau,-\frac{\sqrt{3}}{2}\tau,0 \right)$ & $\left( \mp \frac{1}{2}\tau, \pm \frac{\sqrt{3}}{2}\tau,1 \right)$  \\
		& $\left( -\frac{1}{2}\tau,-\frac{\sqrt{3}}{2}\tau,0 \right)$,$\left( \frac{1}{2}\tau , \frac{\sqrt{3}}{2}\tau , 0 \right)$ & $\left( \pm \frac{1}{2}\tau , \pm\frac{\sqrt{3}}{2}\tau , 1 \right)$ \\
		& & \\
		\hline 
		\hline
		%\bottomrule
	\end{tabular}
\end{table}
%%%%%%%%%%%%%%%%%%%%%%%%%%

%%%%%%%%%%%%%%%%%%%%%%
%%%%%%%%%%%%%%%%%%%%%%
%%%%%%%%%%%%%%%%%%%%%%%%%%%%%%%%%%%%%%%%%%%%%%%%%%%%%%%%%%%%%%%%%%%%%%%%%%%%%%%%%%%%%%%
\section{Experimental REXS Results}

\begin{figure}[h]
	\includegraphics*[width=7cm]{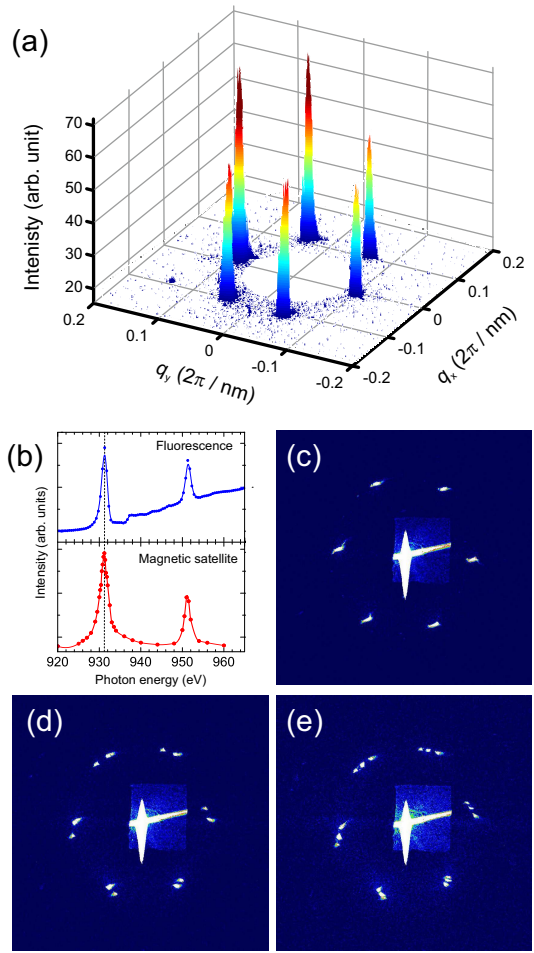}
	\caption{(Color online). (a) Experimentally obtained reciprocal space map in the $hk$ plane around the Cu$_2$OSeO$_3$ (001) Bragg peak (blanked out) in the skyrmion phase. (b) X-ray absorption spectra. Top: fluorescence yield; bottom: integrated satellite intensity. (c-e) CCD images of single, double, and triple split diffraction patterns.}
	\label{fig_3}
	\vspace*{-0.2cm}
\end{figure}

We performed REXS experiments on a well-characterized Cu$_2$OSeO$_3$ single crystal \cite{EPFL-ARTICLE-196610}.
Instead of taking single CCD images at the crystalline (001) Bragg condition, we carried out reciprocal space maps (RSMs) by rocking the sample $\pm 2.5^\circ$ around the (001) peak such that the entire helix propagation-related reciprocal space is covered. Figure \ref{fig_3}(a) shows the RSM of the $hk$ plane ($l=1$) at 56.6~K in an applied field of 30~mT along the (001) direction. The incident x-rays are linearly polarized with a photon energy of 931.25~eV. Six sharp satellite peaks can be observed, corresponding to the skyrmion phase. Moreover, when scanning both temperature and field across the skyrmion phase region, no peak splitting is observed. This suggests a six-fold symmetric equilibrium ordering in the entire skyrmion phase pocket.

%%% FIGURE %%%%%%%%%%%%%

We performed RSMs for each energy point, and plot the spectroscopic profile using the integrated satellite intensity in Fig.\ \ref{fig_3}(b), bottom panel.
Figures \ref{fig_3}(c-e) show single-shot CCD images in the (001) Bragg condition with the field orientation rotated $15^\circ$ away from the (001) direction (in the scattering plane) and from the direction of the incoming x-rays. 
Now, the well-defined six spots split into two, and finally three sets. This confirms the existence of a multidomain skyrmion state, where each domain has a different helix propagation orientation, which can be intentionally created by introducing a magnetic field gradient.
This observation of a multidomain state is consistent with earlier LTEM work by Tokura \textit{et al.}\ \cite{Tokura_CuOSeO-MnSi_ratchet_Natmater_14}, where split peaks (in the Fourier transforms of the LTEM domain patterns) were observed as part of a dynamic domain rotation process (see Supplementary Movie S2 in Ref.~\cite{Tokura_CuOSeO-MnSi_ratchet_Natmater_14}). It has to be noted that x-ray based techniques are sampling a much larger area than electron microscopy based techniques, meaning that a multidomain state observable by LTEM will be picked up by REXS as well.

%%%%%%%%%%%%%%%%%%%%%%
%%%%%%%%%%%%%%%%%%%%%%
%%%%%%%%%%%%%%%%%%%%%%%%%%%%%%%%%%%%%%%%%%%%%%%%%%%%%%%%%%%%%%%%%%%%%%%%%%%%%%%%%%%%%%%
\section{Discussion}

Resonant soft x-ray diffraction experiments on Cu$_2$OSeO$_3$ single crystals has been previously carried out by Langner \textit{et al.} \cite{Langner2014}, where the observation of two sets of six-fold symmetric spots has been reported.
The authors state that the peak splitting arises from the two inequivalent Cu sites. They support this statement by an observed 2-eV difference in energy profiles for the so-called `left' and `right' spot spectra (shown in Fig.~2 of Ref.~\cite{Langner2014}). They further argue that the two-fold splitting is linked to an in-plane rotation of two skyrmion sublattices, leading to a moir\'{e} pattern (Fig.~4 in Ref.~\cite{Langner2014}).

Bond valence sum calculations show that the two inequivalent Cu$^{\mathrm{I}}$ and Cu$^{\mathrm{II}}$ sites in Cu$_2$OSeO$_3$ (where the superscripts I and II refer to the different lattice sites, not to different oxidation states) have practically the same valence charge \cite{Bos2008}, and density functional theory calculations show that their unoccupied states have very similar energies \cite{Yang2012}.
We note that the Cu $L_{2,3}$ transition for Cu$^{2+}$ is $3d^9 \rightarrow 2p^5$$3d^{10}$. In the final state the $3d$ shell is full, which reduces the transition to a one-electron process without $2p$-$3d$ core-hole interaction. This gives a single absorption peak at 931~eV without multiplet splitting \cite{Laan1992}. 
There are no known Cu $d^9$ compounds with such a large splitting energy, and in fact, the energy splitting that could be expected would be well below $\sim$1~eV. As reported by Bos \textit{et al.}\ \cite{Bos2008}, and earlier by other others, Cu$^\mathrm{I}$ and Cu$^\mathrm{II}$ have practically the same valence charge, which can only result in a minute energy shift in the Cu $L_3$ absorption spectrum, well below the energy resolution limit (and certainly less than 2~eV reported in Ref.~\cite{Langner2014}).
Note that the case would be of course very different for systems in which multiplet splitting exists, such as Fe compounds \cite{Gerrit-Theo91}.
Since CuO$_2$ exhibits a peak at 933~eV \cite{Laan1992}, one possible explanation is that the higher energy peak is due to a Cu$^{1+}$ contamination.

Alternatively, another possible source of the discrepancy is the way the energy scans are carried out. 
In particular, spectroscopic data obtained by analyzing the local pixels at the CCD plane is rather inaccurate, as the definition of the `left' and `right' spots on the camera is arbitrary for each energy. Most importantly, for different photon energies, scattering from the same propagation wave vector will result in a shift of the spot in the camera plane, leading to a `rocking-curve-like' Gaussian peak.
Therefore, from a spectroscopic viewpoint, there is no evidence that the split-satellite peak is due to inequivalent Cu sites.

A core argument for the formation of double-split six-fold diffraction patterns, given in Ref.~\cite{Langner2014} (Fig.\ 4), is that the superposition of two in-plane rotated skyrmion sublattices leads to a moir\'{e} pattern.
However, the real-space moir\'{e} pattern and the presented Fourier transform (diffraction pattern) are mathematically not related. Rather, the superposition leads to new components in the Fourier spectrum, in particular six-fold symmetric satellites around the main reciprocal lattice points that result from the long wavelength beating in the moir\'{e} pattern---a phenomenon well-known from hexagonal coincidence lattices \cite{Graphene2014}.

Further, it is important to consider another requirement that has to be met in order to carry out REXS experiments in a quantitative way. In the experiment reported in Ref.~\cite{Langner2014} the skyrmion plane is always perpendicular to the vertical direction of the laboratory reference frame, given by the fixed direction of the magnetic field, and it is thus independent of the goniometer angle $\theta$. There are (at least) six wave vectors ($\tau$) coupled to the (001) Bragg peak, giving rise to the observed magnetic peaks. In Fig.~2(a) in Ref.~\cite{Langner2014}, these magnetic peaks are different in both amplitude and orientation. Therefore, for a single $\theta$, it is impossible to reach the diffraction conditions for all six magnetic peaks at the same time.
The observed skyrmion pattern (CCD images in Figs.~1(b) and 2(a) in \cite{Langner2014}) was collected in the \textit{structural} (001) Bragg condition, which is not the correct diffraction condition for either of the magnetic satellites. As a result of this, the satellites still have intensity, analogously to sitting at the edge of a rocking-curve peak.
A single-shot CCD image corresponds to a curved plane in reciprocal space, which is not equal to the skyrmion plane in reciprocal space. As a result of this, the skyrmion diffraction spots will not end up on a circle, but on an oval, as can be seen in Fig.~2(a) in Ref.~\cite{Langner2014}.
These satellite spot on the camera does not correspond to the peak position of (001)+$\tau$, but a poorly-defined reciprocal space point that could largely deviate from (001)+$\tau$. Also, the magnetic peaks of (001)+$\tau$ will not necessarily appear on the same oval for a single goniometer angle as they do not reach the diffraction conditions at this angle.

Instead, a much more simple explanation of a peak splitting in this context is the occurrence of two \textit{non-superimposed} skyrmion lattice domains that are simultaneously sampled by the wide x-ray beam.

%%%%%%%%%%%%%%%%%%%%%%
%%%%%%%%%%%%%%%%%%%%%%
%%%%%%%%%%%%%%%%%%%%%%%%%%%%%%%%%%%%%%%%%%%%%%%%%%%%%%%%%%%%%%%%%%%%%%%%%%%%%%%%%%%%%%%
\section{Summary and Conclusions}

In conclusion, we used REXS on the chiral magnet Cu$_{2}$OSeO$_{3}$. We presented a detailed discussion of the magnetic contrast stemming from the magnetic phases. We showed experimental results of the six-fold symmetric magnetic diffraction pattern, in which the peaks were unsplit, double-split, as well as triple-split, depending on the magnetic history of the sample. This clearly contradicts the interpretation given in Ref.\ \cite{Langner2014} where the double-split peaks have been associated with the two chemically distinct Cu sites. Instead, by carefully performing XAS measurements, we find no evidence of a peak splitting. Oppositely, a more simple explanation is the occurrence of a multidomain skyrmion state, sampled by the relatively wide x-ray beam.

%%%%%%%%%%%%%%%%%%%%%%
%%%%%%%%%%%%%%%%%%%%%%
%%%%%%%%%%%%%%%%%%%%%%%%%%%%%%%%%%%%%%%%%%%%%%%%%%%%%%%%%%%%%%%%%%%%%%%%%%%%%%%%%%%%%%%
\section*{Acknowledgments}
The REXS experiments were carried out on beamline I10 at the Diamond Light Source, UK, under proposals SI-11784 and SI-12958.
S.~L.~Z.\ and T.~H.\ acknowledge financial support by the Semiconductor Research Corporation.
A.~B.\ and C.~P.\ acknowledge financial support through DFG TRR80 and ERC AdG (291079, TOPFIT).

\end{document}